# Computational International Relations

What Can Programming, Coding and Internet Research Do for the Discipline?

**Author**: H. Akin Unver, assistant professor of international relations, Kadir Has University, and a dual non-resident research fellow at the Center for Technology and Global Affairs, Oxford University and the Alan Turing Institute in London. akin.unver@khas.edu.tr

**Abstract**: Computational Social Science emerged as a highly technical and popular discipline in the last few years, owing to the substantial advances in communication technology and daily production of vast quantities of personal data. As per capita data production significantly increased in the last decade, both in terms of its size (bytes) as well as its detail (heartrate monitors, internet-connected appliances, smartphones), social scientists' ability to extract meaningful social, political and demographic information from digital data also increased. A vast methodological gap exists in 'computational international relations', which refers to the use of one or a combination of tools such as data mining, natural language processing, automated text analysis, web scraping, geospatial analysis and machine learning to provide larger and better organized data to test more advanced theories of IR. After providing an overview of the potentials of computational IR and how an IR scholar can establish technical proficiency in computer science (such as starting with Python, R, QGis, ArcGis or Github), this paper will focus on some of the author's works in providing an idea for IR students on how to think about computational IR. The paper argues that computational methods transcend the methodological schism between qualitative and quantitative approaches and form a solid foundation in building truly multi-method research design.

# Introduction –
## What is Computational IR (*ComInt*)?

Computational International Relations (ComInt[1]), introduced as a specific inquiry of research in this paper, derives from the Computational Social Science (ComSoc[2]) revolution of the last decade. International relations (IR) literature has long trailed behind political science (PolSci) since the seminal Designing Social Inquiry[3] (DSI) by King, Keohane and Verba. Setting the quantitative bounds of the discipline despite its evolution over the years, DSI has established the methodological orthodoxy of both IR and PolSci, becoming the key text in almost all methods classes. The showdown of critical and supportive camps over DSI has continued well into today, setting the parameters of methodological polarization. The strong empiricism of regression and statistical modelling was challenged by the qualitative camp for a variety of reasons, including distortion of analytical focus[4], manipulation of data[5], and overall skepticism over how much mathematical validity can imply causality[6]. This long and seemingly unending core debate on methodology in IR and PolSci has become eclipsed by the advent of computational social science as a meta-bridge between extreme ends of hard sciences and social sciences.

There is no one single gateway to computational social science. It is rather a meeting point between diverse disciplines that seek to strengthen their analytical approach through the use of a wide array of computer and data science tools. Although the term used to define computer (or other hard) scientists using big data processing methods to explain social phenomena[7], this frame is currently expanding. Increasingly more social scientists are getting trained in the ways of data science and Internet research, harvesting new forms of data to expand some of the fundamental assertions of their literature. Training a dedicated 'computational social scientist' is a complicated and broad task, with many definitional and operational questions. For example, how can a student with no programming or computer background start learning computational tools? Which exact programming languages should a student master? Is it enough to learn coding? How much? Once coding yields promising data, should you map it, or run a cluster network analysis? Should you learn Python or R? Is it better to specialize in geospatial research, sentiment analysis, neural networks or data mining? The difficulty of answering these questions, beyond the fact that such answers are highly subjective, lies within the rapidly transforming technical environment of computer science.

*A Brief History of Computational Social Science (ComSoc)*

It is hard to build an accurate trajectory of ComSoc. Different social sciences disciplines have adopted, dropped, marginalized and re-adopted computer-based tools at different points since 1980s. Earliest forms of Dynamic Systems Theory[8] and Artificial

---

[1] I will be abbreviating Computational IR as ComInt, as CIR is an over-crowded abbreviation in international relations, used in reference to Coordinator for International Relations, Central Intelligence Report, Critical Information Requirements, CIR Capital Investments Review, among many others.

[2] I am strongly in favour of abbreviating Computational Social Science as ComSoc, because CSS is already over-crowded with several computer science-related terminologies (including 'computer science'), including Cascading Style Sheets, Content Scrambling System, Central/Computing Support Services, Core System Software, Client Security Software and so on…

[3] Gary King, Robert O. Keohane, and Sidney Verba, *Designing Social Inquiry: Scientific Inference in Qualitative Research* (Princeton, N.J: Princeton University Press, 1994).

[4] Christopher H. Achen, "Let's Put Garbage-Can Regressions and Garbage-Can Probits Where They Belong," *Conflict Management and Peace Science* 22, no. 4 (September 1, 2005): 327–39, https://doi.org/10.1080/07388940500339167; David Collier and Henry E. Brady, *Rethinking Social Inquiry: Diverse Tools, Shared Standards* (Lanham, Md: Rowman & Littlefield Publishers, 2004); Philip A. Schrodt, "Beyond the Linear Frequentist Orthodoxy," *Political Analysis* 14, no. 3 (July 2006): 335–39, https://doi.org/10.1093/pan/mpj013.

[5] Frank P. Harvey, "Practicing Coercion: Revisiting Successes and Failures Using Boolean Logic and Comparative Methods," *The Journal of Conflict Resolution* 43, no. 6 (1999): 840–71; Peter A. Hall, "Aligning Ontology and Methodology in Comparative Research," in *Comparative Historical Analysis in the Social Sciences*, ed. James Mahoney and Dietrich Rueschemeyer, Cambridge Studies in Comparative Politics (Cambridge: Cambridge University Press, 2003), 373–404

[6] John Gerring, *Social Science Methodology: A Criterial Framework* (Cambridge; New York: Cambridge University Press, 2012); Judea Pearl, *Causality: Models, Reasoning and Inference*, 2nd edition (Cambridge, U.K. ; New York: Cambridge University Press, 2009).

[7] Steven Bankes, Robert Lempert, and Steven Popper, "Making Computational Social Science Effective: Epistemology, Methodology, and Technology," *Social Science Computer Review* 20, no. 4 (November 1, 2002): 377–88, https://doi.org/10.1177/089443902237317; Cristiano Castelfranchi, "The Theory of Social Functions: Challenges for Computational Social Science and Multi-Agent Learning," *Cognitive Systems Research* 2, no. 1 (April 1, 2001): 5–38, https://doi.org/10.1016/S1389-0417(01)00013-4; Flaminio Squazzoni, "A (Computational) Social Science Perspective on Societal Transitions," *Computational and Mathematical Organization Theory* 14, no. 4 (December 1, 2008): 266–82, https://doi.org/10.1007/s10588-008-9038-y.

[8] Walter C. Hurty, "Dynamic Analysis of Structural Systems Using Component Modes," *AIAA Journal* 3, no. 4 (1965): 678–85, https://doi.org/10.2514/3.2947; Erich Jantsch, "From Forecasting and Planning to Policy Sciences," *Policy Sciences* 1, no. 1 (March 1, 1970): 31–47, https://doi.org/10.1007/BF00145191; J Brian McLoughlin and Judith N Webster, "Cybernetic and General-System Approaches to Urban and Regional Research: A Review of the Literature," *Environment and Planning A* 2, no. 4

Intelligence[9] debates of the 1950-60s have led to the emergence of Complexity Science[10] and the popularization of Agent-Based Modelling[11] in sociology and behavioral economics. As computers became more powerful and widely available in the 1980s, first forms of Data Mining[12], Genetic Algorithms[13] and System Dynamics Models[14] emerged in social research. Through the 1990s, earlier adoptions of Internet data on social research began to emerge, creating multi-layered connections into complexity research, network science and urban systems modelling. At certain times, these attempts merged into the existing quantitative strand in social sciences, where in others, computational progress embarked on its own journey, steering clear of mainstream statistical and mathematical methods. By 2000s, computer models of large sets of quantified data were already being used in cognition, decision-making, behavioral approaches, groups and organization, social interactions and systemic analysis of world events. Conte et. al.[15] identify three main schools of development in ComSoc: deductive (macro theory-building through mathematical modelling and computer processing), generative (micro theory-building through behavioral modelling and computer simulations) and complexity science (use of non-static large and live datasets to explain and forecast behavior and choice-based uncertainty) variants.

With the emergence of digital platforms and social media, and global proliferation of smartphones, ComSoc departed from its previous focus and began harvesting this new, abundant and highly granular type of digital data. Current definitions of ComSoc therefore distinguish between computer-based social science[16], which is using computer programs to process quantitative social data and ComSoc, which processes enormous chunks of - often real-time - Internet data[17]. Although the quantity and granularity of digital data produced every day is impressive, a key question remains how to process such data in a meaningful way and how to build social theory using it. As of July 2016, Instagram, Twitter, Facebook and other social media platforms combined, produced around 650 million publicly available posts per day[18], making up '*the largest increase in the expressive capacity of humanity in the history of the world*'[19]. With the emergence of increasingly more powerful computers, along with most creative data processing software, all scientific disciplines gained access to historically unprecedented and unfathomably detailed information on micro and macro-level human interactions.

Computational IR (hereafter, *ComInt*) derives largely from the founding and advent of ComSoc. in the last few years. Related to, but separate from ComSoc, ComInt deals exclusively with core IR topics of power, conflict/peace, state behavior, international norms/institutions and the world system/order. As ComInt starts dealing with non-state actors (NGOs, MNCs, media, religious groups, Diasporas, militants etc.) it steers further into the domain of sociology and shares common ground with digital, or tech sociologists. This domain requires even further novel methods, as tracing the transient shifts and trends of non-state actors require a way to bring ethnography close to the field of computational methods that both include, but also expand upon the existing approaches of digital and/or Internet ethnography.

Both data scientists and natural sciences scholars I got the luck of working with at Oxford Internet Institute, Oxford Computer Science Department and the Alan Turing Institute had a distinct interest in the realist strand of IR. They had an automatic tendency to accept states as singular and primary units of analysis in their approaches and without exception, all of them wanted to address questions related to survival, conflict and security - all from a state-centric point of view. Defense, balance of power, armed conflict and resource-infrastructure (capability) oriented research agendas have attracted significantly more computational research attention than other promising approaches in IR such as constructivism, post-structuralism, critical or post-modern theories. This is a shame, as I will demonstrate later on, data mining, entity recognition or geo-statistical mapping methods can successfully challenge a number of these approaches.

Defined in simple terms ComInt, relies on the mining and processing of vast quantities of digital social footprint to study, model and explain world events. In doing that, it transcends the traditional schism between qualitative and quantitative methodology and presents a 'third way' methodology that frees the researcher from the restrictions of both methodological schools. ComInt predominantly (but not exclusively) uses large chunks of digital footprint and focuses on social online activities that generate enormous quantities of social data. This is one of the reasons why ComInt or ComSoc didn't exist a decade ago, and also a reason why merely using numerical analysis software like R, Python and MatLab to model existing quantitative data, isn't really ComSoc or ComInt. The origin, size and type of data that is collected make the main difference, as well as the main focus of study; the Internet and digital interdependencies.

*Charting the Waters: Main Questions of ComInt*

ComInt is an emerging field with yet unclear analytical borders. How to study, research and teach it remains a developing endeavour. The purpose of this section is to set five main signal posts across a vast scientific territory, helping newcomers to identify and scale out the field. These five signal posts constitute five different approaches to digital data processing, along with their theoretical and research design point of view: a) language/text, b) mapping, c) modelling, d) communication and e) networks.

*i.     Language and Text*

Although linguistics has so far predominantly been used by the qualitative part of social sciences and IR, computational tools introduced new approaches to the quantitative study of large amounts of text. Quantitative linguistics[20] existed as a vibrant sub-discipline as far back in the 1960s, but the advent of computational linguistics[21] as relevant to IR is a relatively new phenomenon, since it renders historically unprecedented volumes qualitative information usable by text processing programs. Traditionally, IR's relationship with linguistics has largely been driven through critical discourse analysis, where qualitative data was interpreted as 'unmeasurable'[22], containing qualities like emotion, sentiment and judgement that couldn't conceivably be analyzed through numeration. Digitization of text and the advent of speech recognition technologies have enabled large chunks of text to be searchable. Then came the process by which vast quantities of parliamentary archives, historical documents and official statements became digitized, bringing text analysis into the domain of computation. Today, thanks to the Internet and social media, more than 7 million web pages[23] of text are being added to our collective repository of text, searchable, quantifiable and measureable.

Text mining tools such as WordStat, RapidMiner, KHCoder aim to dig into vast quantities of written resources and even real-time transcribed speech through specialized computer software. They differ fundamentally from online text searching tools such as

---

[20] Gustav Herdan, "Quantitative Linguistics or Generative Grammar?," *Linguistics* 2, no. 4 (2009): 56–65, https://doi.org/10.1515/ling.1964.2.4.56.
[21] Kenneth W. Church and Robert L. Mercer, "Introduction to the Special Issue on Computational Linguistics Using Large Corpora," *Computational Linguistics* 19, no. 1 (March 1993): 1–24.
[22] Richard E. Palmer, "Postmodernity and Hermeneutics," *Boundary 2* 5, no. 2 (1977): 363–94, https://doi.org/10.2307/302200; Paul Hernadi, "Dual Perspective: Free Indirect Discourse and Related Techniques," *Comparative Literature* 24, no. 1 (1972): 32–43, https://doi.org/10.2307/1769380.
[23] http://www.internetlivestats.com/

Google or Bing by allowing the analyst to establish connections, detect patterns and build networks between very large text datasets, rather than merely searching within them. Although these tools and methods are being advanced and updated at great pace, some analytical elements remain recurrent across most studies.

- *Information Retrieval* is perhaps the oldest of text mining tools and one that is the easiest to replicate by simple programming. Information retrieval identifies the documents in a text dataset to match a specific search term. Google, Yahoo or Bing search engines are the best-known information retrieval systems, and almost all libraries use a version of these systems.
- *Natural Language Processing* on the other hand bring text mining into the domain of artificial intelligence: how can computers understand a diverse set of human language in a way that humans communicate with each other? In other words, how can a computer automatically identify verbs, nouns, emotions, threats and sarcasm in a new language it is introduced to? How should an algorithm recognize 'Donald Trump is a great President' when the statement is used as sarcasm, as part of a critical tweet, for example? Natural Language Processing is crucial to 'teach' a computer how to code and interpret different, hidden meanings in language, as well as culturally-contingent, unique expressions. This process is usually the first step in building a corpus (body of textual digital knowledge) to build information extraction and data mining systems.
- *Information Extraction* is the process by which unstructured textual data is reorganized into a structured form based on the corpus obtained by Natural Language Processing. It is further split into three main approaches
    - Term Analysis, which extracts different versions and references to the same term in a range documents, especially when these documents include a mixture of official, unofficial, translated and native-language versions.
    - Named-Entity Recognition, which extracts people, organizations or groups, in addition to different expressions of numbers (percentages, time and location)
    - Fact Extraction, which identifies and extracts relationships, networks and subtle connections in a document, such as between entities, events, dates and geographic designations.

The resultant processes enable the researcher to discover previously unidentified and unestablished knowledge from text and especially by studying large bodies of text in relation to each other. For example, by deep diving into American, British, Russian and French official documents about the 1945 Yalta Conference might give us comparative information over how all four sides understood and interpreted the terms of the conference, allowing us to generate new knowledge in diplomatic history over how these countries structured their foreign policies through early Cold War.

Some of the most promising employments of computational language and text analysis on international relations involve sentiment analysis, detection of certain types of behavior (such as radicalization) through text, opinion mining and violent event detection/prediction. In one of the most relevant cases for IR, Bermingham et al.[24] demonstrate how harvesting word and sentiment combinations in Youtube's comment section of jihadi videos can offer a predictive model of radicalization. Delving beyond the scope of the Internet, Hsinchun Chen[25] for example, has discovered an automated sentiment mining model the Dark Web text data, presenting a methodological avenue for the detection of potential radicalization online. Dubvey et. al. took out automated sentiment mining methods beyond radicalization/terrorism research and harvested digital media posts of Indian diplomats. In doing so the researchers presented a promising model on how foreign service members interact with politics in online space. Finally, a personal favorite of mine, Hannes Mueller and Christopher Rauh have recently published an excellent paper[26], which successfully predicts political violence by harvesting automated newspaper text. The authors argue that it is possible to predict armed conflict and political violence in a specific country by analyzing within-country variation of topics in national newspapers. Finally, one of the earliest forms of IR-relevant text mining studies in Turkey has been conducted by Hatipoglu et. al.[27] on how digital information and content diffusion on 2015 'Kobani riots' in Turkey influenced foreign policy perceptions at the national scale.

---

[24] Adam Bermingham et al., "Combining Social Network Analysis and Sentiment Analysis to Explore the Potential for Online Radicalisation," in *Proceedings of the 2009 International Conference on Advances in Social Network Analysis and Mining*, ASONAM '09 (Washington, DC, USA: IEEE Computer Society, 2009), 231–236, https://doi.org/10.1109/ASONAM.2009.31.

[25] H. Chen, "Sentiment and Affect Analysis of Dark Web Forums: Measuring Radicalization on the Internet," in *2008 IEEE International Conference on Intelligence and Security Informatics*, 2008, 104–9, https://doi.org/10.1109/ISI.2008.4565038.

[26] Hannes Mueller and Christopher Rauh, "Reading Between the Lines: Prediction of Political Violence Using Newspaper Text," *American Political Science Review*, December 2017, 1–18, https://doi.org/10.1017/S0003055417000570.

Text mining is useful when it is used alongside another method, such as discourse analysis, process tracing or statistics. Furthermore, text mining research groups perform better when they contain a subject expert (historian, sociologist or ethnographer) and a linguist (theorist) alongside programmer(s) and engineer(s), in contrast to research clusters that contain only the latter two. Usually the hardest and unfortunately the most overlooked aspect of text mining is building a corpus that is culturally, contextually and case-wise aware of the nuances and subtleties of the language(s) that is/are being studied. Furthermore, both the corpus and search terms need to be grounded in theory, in order to avoid concept stretching or build a corpus with redundant or irrelevant terms.

*ii.    Mapping*

Mapping and geospatial analysis contribute to some of the most central components of IR, including geopolitics/geography, borders and space. It is also one of the most popular approaches to generating event data, which allows researchers to display spatial dynamics of war, conflict and inequality. The terms GIS (Geographic Information Systems) and 'geospatial' are usually used interchangeably, with often unclear differences that separate the two. Both approaches refer to visual systems where geographic information is stored in layers, that are then viewed, manipulated and measured through a dedicated mapping software. The essence of mapping research is geographical coordinates and other geographical data (altitude, topography, elevation, depth, etc.) that are often coupled and analyzed in relation to tabular data that contains various sets of statistical information such as landmarks, infrastructure or econometric data.

Mapping has become exceptionally relevant and important to the study of IR and other social sciences, mainly because of the introduction of geolocation information integrated into smart devices and social media. Through the generation of large sets of social data containing geographical information, researchers are now able to study social and political phenomena with much higher level of granularity, sometimes in real-time. Instead of using geographical data to study natural resources, transportation infrastructure or household income, we are increasingly able to derive behavioral information through micro expressions of digital activity. (Foursquare check-ins, Facebook status updates, tweets containing geo-location information) This yields vast quantities of new information related to human behavior and relations for the use of emergency response teams (emergency behavior of large groups of people), local governments (transportation and traffic behavior of individuals), companies (purchasing power analysis, response to advertisements, marketing analysis), among others.

Geospatial research is conducted both on dedicated GIS application packages (such as ArcGis, QGis), or mainstream data processing-programming platforms that have GIS plug-ins (like Python and R). Even Excel is experimenting with mapping plug-ins that can be used with existing .xls or .csv files. Geospatial data on the other hand, is divided into two categories: vector and raster data. Vector data refers to points and polygons that designate or enclose a specific coordinate on a base map, such as coordinates of schools in a geographical area, or allocated farmland in a rural province. Raster data on the other hand refer to aerial imagery and digital elevation models that render a map three dimensional. While raster data is not really necessary to analyze school districts, it is crucial for the study of river flows or transportation systems. These datasets are usually stored in dedicated geodatabases that can be downloaded for study, or users can generate their own datasets through manual entries, or web scraping techniques. Increasingly, LiDAR (Light Detection and Ranging), UAVs (unmanned aerial vehicles), GPS (geographic positioning systems) and satellites have begun to be used more frequently to generate open-access geospatial analysis data.

Some of the best examples of geospatial IR excel not only in finesse in visualizing location data, but successfully tell a story that builds and tests a theory. Hein E. Goemans and Kenneth A. Schultz for example, demonstrated how states in Africa make claims to some border areas and not others through aggregating a digital geospatial dataset of border disputes in a cross-continent study.[28] The ingenuity of the piece is that it discovers territorial contestation taking root not from natural boundaries such as watersheds or rivers, but mostly from historical-colonial contestation points. Mark Graham et al. on the other hand, have demonstrated how digital labor affects global worker micro-economies, specifically in terms of how type of online labor influences worker bargaining power, economic inclusion and worker livelihoods.[29] By using a dataset showing geographic engagement with digital labor, the researchers come up with both micro-level behavior and macro-level structures of what could be termed as 'digital Marxist research'. In one of the most famous examples of geospatial IR dataset construction, Alberto Alesina et. al. delve into the origins of ethnic inequality using satellite-generated nighttime luminosity data.[30] By exploring time-frequency distribution of electricity, the researchers come up with important tests of intra-state ethnic inequality theories, including more macro-scale developmental and economic disparities within and across countries. One of the most interesting newer studies on geospatial proximity networks has been conducted by Jesse Hammond[31], who demonstrated that network of roads and connections between population centers are the primary determinants of conflict onset and diffusion in civil wars. This study challenges previous findings on geography and conflict by discovering a significant reporting bias in the building of past conflict datasets.

---

[27] Emre Hatipoğlu et al., "Sosyal Medya ve Türk Dış Politikası: Kobani Tweetleri Üzerinden Türk Dış Politikası Algısı," *Uluslararası İlişkiler* 13, no. 52 (2016): 175–89.
[28] Hein E. Goemans and Kenneth A. Schultz, "The Politics of Territorial Claims: A Geospatial Approach Applied to Africa," *International Organization* 71, no. 1 (January 2017): 31–64, https://doi.org/10.1017/S0020818316000254.
[29] Mark Graham, Isis Hjorth, and Vili Lehdonvirta, "Digital Labour and Development: Impacts of Global Digital Labour Platforms and the Gig Economy on Worker Livelihoods," *Transfer: European Review of Labour and Research* 23, no. 2 (May 1, 2017): 135–62, https://doi.org/10.1177/1024258916687250.

*iii.    Modelling*

Mathematical and physical modelling of social phenomena aren't new. Since 1960s, applying natural sciences principles and functions on social events have been amply used by researchers.[32] The advent of computational methods allowed computer science to bridge this interdisciplinary gap between natural and social sciences. In the last decade, three types of main modelling approaches have grown in popularity: mathematical, physics-based and bio-organistic models. These approaches allow us to better study mass social events like voting, riots, war and political engagement at an unprecedented granularity. The advent of computational methods significantly increased the impact and relevance of all three modelling approaches for social sciences and IR.

Mathematical modelling of social phenomena is structured upon a somewhat problematic assumption that human behavior can be observed within and based on arbitrarily set constants and can be measured in numerical terms. Although elements such as uncertainty, chance and bias are added into models, the foundational assumption that human behavior can be quantified is still there.[33] While this assumption is subject to a separate set of epistemological debates, mathematical models of human and social behavior are nonetheless both popular and useful in testing concepts such as equilibrium/non-equilibrium, stability/instability or order/chaos that can assume subjective meanings without proper measurement. Mathematical models in social sciences are structured upon a number of sub-approaches such as voting/preference (Arrow's impossibility theorem[34], Shapley-Shubik index[35]), dynamic models (Richardson arms race model[36], Lanchester combat models[37], predator/prey model[38]) and ecology (phase space[39], boxicity[40]) and stochaistic models (Markov chains[41], learning theory[42], social power approach[43]).

Physics models are more complex and less intuitive for social sciences, mainly because of lack of bridging literature between physics and social sciences. Although the number of approaches are growing by the emergence of a new breed of interdisciplinary researcher doing this bridging work, there are roughly two main types of physics modelling that can meaningfully be adapted into social sciences. The first of those, cellular automata[44] for example, deals with the interaction of particle systems in parallel and sequential dimensions. Take for example the spread of war and conflict between neighboring countries. Each country *i* becomes 'infected' with war *(Si = 1)*, if at least one of its nearest neighbors is already witnessing conflict. Computational tools handle this diffusion mechanism well, predicting and modeling the likelihood of, for example, infection *(i)* to be spread across 24 different neighboring countries in time *t*, coming up with a predictive model of how far and how fast can the conflict spread into adjacent territories. The second type of physics modelling derives from temperature models (Boltzmann probability[45]). They give us how energy and pressure are diffused across different units and how systems enter into entropy or recalibration based on the latent or free energy travelling between the constituent elements of the system. To clarify for our purposes, Boltzmann probability would give us diplomatic pressure, international bandwagoning or buck-passing behavior within an alliance or regional system: if your allies sign a diplomatic treaty, they can also influence you into signing the same treaty, even though the said treaty may not be in your country's interests. Thus, let *E* be the number of closest allies signing the treaty, minus the number of allies that aren't signing the treaty. The probability for your country to switch then is given by the energy different and equal to *exp(−2E/T) (or 1 if E < 0)*, generating the terms in which you can withstand the pressure from treaty-signing allies and refrain from signing the treaty that doesn't serve your national interests. Both cases can be adopted into ComInt through testing theories on voting in international institutions, alliance behavior, international financial markets and interactions between security cultures.

Bio-organistic types of modelling also substantially derive from mathematical and physics modelling. However, one particular type of biology model penetrated more than other types into the domain of social sciences: epidemiology[46]. Epidemiological modelling is a simplified version of describing transmission of diseases through a pre-determined network of agents. Epidemiological models allow social scientists to make sense of collective action and large-scale popular mobilization in the form of riots, protests, migration and emergency social behavior, such as disasters. Epidemiological models based on mathematical formulations of how infectious diseases spread, can offer to make sense of complex social behavior that would otherwise be very hard to monitor and measure. There have been different methodological approaches to the study of complex social behavior such as agent-based models, spatial data studies and simple mathematical formulations. What makes epidemiological models different from past methods is its conceptualization and modes of measurement on disorder, uncertainty and unpredictable complexity.

One of the most fascinating and novel studies on social epidemiology has been Laurent Bonnasse-Gahot et. al.[47] seminal study on how 2005 French riots spread and were contained. Building a riot contagion model, the authors assess geographic proximity, social networks and riot outcome in explaining how neighborhood/district relations have been instrumental in the diffusion of these riots. Such riot and social movement modelling works are of direct interest for IR scholars as they will substantially strengthen some of the existing IR and PolSci theories on how conflicts start, diffuse and end. A second key study is Toby Davies et. al.[48] account of how 2011 London riots and their policing have followed a direct spatial contagion model, building a high-granularity digital event dataset. The researchers test a number of IR-relevant topics such as force deterrence, local escalation models and crisis signaling, through measuring police-to-riot distance, with the added variables of police versus rioter numbers. Finally, both Guo et. al.[49] and Kirby and Ward[50] make attempts to generate a macro explanation of war and peace through spatial modelling. Guo et. al. formulate 'betweenness centrality' (a physics principle) in order to assert that cities that have the highest betweenness factor (population density, ethnic fractionalization versus the number of outside connections) are more likely to contain conflicts in geographies in-between. Kirby and Ward on the other hand reject nation states as the primary actors in peace and war, and using a digital dataset from Africa, they argue that it is the local and tribal relations that determine the course and extent of state-level violence.

### iv. Communication

Digital technologies have brought forward another big leap in communication, comparable to the effect of the invention of writing, telegram and telephone. Thanks to digital technologies, we communicate more frequently, in verbal and non-verbal ways (such as emojis, or 'like's) allowing us to engage with a multitude of social, political and economic activities simultaneously. The rise of social media too, has allowed us to view and measure human communication in interactive and forum-like settings, leading to the testing of central IR communication topics like misinformation, uncertainty, signaling and cognitive bias.[51] Furthermore, Internet and social media have fundamentally changed how we seek information, access the news and form our opinion on political and social matters.[52] An added factor is how social media algorithms are acting as intermediaries in our political searches, giving us non-random search results based on a number of parameters.[53] This means that how people access and consume facts and information online may be different than another, leading to sustenance or exacerbation of polarization in political views.[54] The issue of how political information is communicated online and represented in digital news media has become a key debate in political science and one that has significant implications for IR. How do key foreign policy actors and decision-makers use social media? How does the Internet facilitate or impede information-seeking behavior of citizens and politicians during an international crisis? How does different consumption patterns of digital news influence how citizens and politicians view and understand diplomacy and in turn, how does these patterns translate into actual foreign policy?

A newly emerging field of research in digital communication is the advent of bots (automated accounts) in digital space, fueling fake news and misleading information that exacerbate international crises and often lead to popular unrest. Fake news is conceptualized as misleading, incomplete or out of place information that is deliberately directed towards consuming and distraction online attention.[55] Although fake news can be driven by human accounts, recent scholarly attention has focused on how automated accounts (bots) help distribute such news during crucial time frames, such as pre-election periods or international crises.[56] Bot research has thus suddenly become a key topic in political science and concerns IR directly, although the subject of inquiry sits at the intersection of computer science and communication theory.

From a methodological standpoint, Derek Ruths and Jürgen Pfeffer have already demonstrated[57] how social media – although not always representative[58] – can offer better results compared to traditional polling. This is both due to significant biases associated with social media access and expression, but also the blurry picture provided by bots. Kollanyi et. al. has demonstrated[59] how bots have influenced the results of the US Presidential elections; a study that was repeated in Forelle et. al. work[60] on bots during Venezuelan elections. Although this seems like a political science question, external involvement and disruption in national elections is definitely a problem for international relations, explained in detail in Taylor Owen's book on how digital disruption is contextualized in IR.[61] An especially vibrant debate currently revolves around Russian capabilities as a 'bot superpower', able to disrupt and distract political processes in Western countries.[62] Further detailed accounts of Chinese governmental controls on social media and what it means for state-society relations have been beautifully modelled in King et. al. 2013[63] and 2017[64].

---

[51] Bruce Bimber, "Information and Political Engagement in America: The Search for Effects of Information Technology at the Individual Level," *Political Research Quarterly* 54, no. 1 (March 1, 2001): 53–67, https://doi.org/10.1177/106591290105400103.

[52] A great overview of the primary debates in this field can be found in: Andrew Chadwick and Philip N. Howard, *Routledge Handbook of Internet Politics* (Taylor & Francis, 2010).

[53] Pablo Barberá et al., "Tweeting From Left to Right: Is Online Political Communication More Than an Echo Chamber?," *Psychological Science* 26, no. 10 (October 1, 2015): 1531–42, https://doi.org/10.1177/0956797615594620; Helen Nissenbaum Lucas D. Introna, "Shaping the Web: Why the Politics of Search Engines Matters," *The Information Society* 16, no. 3 (July 1, 2000): 169–85, https://doi.org/10.1080/01972240050133634.

[54] Markus Prior, "Media and Political Polarization," *Annual Review of Political Science* 16, no. 1 (2013): 101–27, https://doi.org/10.1146/annurev-polisci-100711-135242.

[55] Nic Newman et al., "Reuters Institute Digital News Report 2017," SSRN Scholarly Paper (Rochester, NY: Social Science Research Network, June 1, 2017), https://papers.ssrn.com/abstract=3026082.

[56] Samuel C. Woolley, "Automating Power: Social Bot Interference in Global Politics," *First Monday* 21, no. 4 (March 10, 2016), https://doi.org/10.5210/fm.v21i4.6161; Samuel C. Woolley and Philip N. Howard, "Automation, Algorithms, and Politics| Political Communication, Computational Propaganda, and Autonomous Agents — Introduction," *International Journal of Communication* 10, no. 0 (October 12, 2016): 9.

[57] Ruths and Pfeffer, "Social Media for Large Studies of Behavior."

[58] Jonathan Mellon and Christopher Prosser, "Twitter and Facebook Are Not Representative of the General Population: Political Attitudes and Demographics of British Social Media Users," *Research & Politics* 4, no. 3 (July 1, 2017): 2053168017720008, https://doi.org/10.1177/2053168017720008.

[59] Bence Kollanyi, Philip N. Howard, and Samuel C. Woolley, "Bots and Automation over Twitter during the First U.S. Presidential Debate," Data Memo (Oxford, UK: Oxford Internet Institute, October 2016), https://assets.documentcloud.org/documents/3144967/Trump-Clinton-Bots-Data.pdf.

[60] Michelle Forelle et al., "Political Bots and the Manipulation of Public Opinion in Venezuela," SSRN Scholarly Paper (Rochester, NY: Social Science Research Network, July 25, 2015), https://papers.ssrn.com/abstract=2635800.

[61] Taylor Owen, *Disruptive Power: The Crisis of the State in the Digital Age*, 1 edition (Oxford ; New York: Oxford University Press, 2015).

A secondary strand of IR-related literature in digital communication is the extent to which online campaigning affects political processes and social mobilization. Koc-Michalska et. al.[65] has explained how online campaigning affects political elections in France, Germany, Poland and the UK, demonstrating that resources, rather than innovation determines success in digital campaigns. In a similarly pessimistic study, Margetts et. al.[66] ran an experiment, testing how political information online affects decision-making of individuals, finding that online information is behavior-changing when it is shared by large groups of people. If the information – whether true or false – isn't shared by a critical mass (or 'social network capital'[67]), it has little influence over political behavior, the study finds. In an interesting twist, Yasseri and Bright explore digital information seeking behavior through Wikipedia traffic data, discovering that political parties whose Wikipedia pages witness a surge in visits close to elections, tend to do well in those elections, compared to other candidates or parties that haven't enjoyed Wikipedia attention.[68] This model can be replicated in to study UN voting patterns or elections within international organizations.

*v.     Networks*

Digital network research is another field that is growing in popularity and allows researchers to study political and power relations in digital space. Often, creative computational researchers discover digital relations and influence maps that cannot be discovered through research in physical space - either due to the controversial nature of the topic, or the difficulty in finding data. Extremism and radicalization networks are the primary foci of computational network analysis. Through digital relations, researchers are able to find influencers, hierarchies and relations in digital space. This could be employed to discover diplomatic networks at the state and institutional level, as well as networks of radicalization at the non-state and sub-state actor level. Ideology research too, can benefit greatly through computational methods, by the use of entity extraction algorithms.

Classical network theory[69] focuses on social networks among individuals (friendships, advice-seeking..) and formal contractual relationships (alliances, trade, security community). What makes network theory important to social science, politics and IR is its ability to conceptualize and theorize relations at the micro, meso and macro-levels of analysis in political processes, offering a structure to seemingly complex interactions.[70] Accordingly, network theory stipulates that relations and internal-external pressures on those relations have the ability to affect beliefs and behaviors.[71] Instead of adopting IR's mainstream levels of analysis approach, network theory focuses on the interactions between these levels of analyses, aiming to conceptualize how these interactions lead to policy and behavior.[72] Computational network analysis on the other hand, takes classical network theory to vast levels of size and complexity, not only designating relations between them, but also use artificial intelligence, machine learning and neural networks approaches to automatically generate real-time changes in these relations.[73]

One of the most relevant recent complex network studies to IR is Jonathan Bright's work on identifying online extremist networks and the role of ideology in polarized digital structures.[74] This work is relevant to IR, because it covers around 90 different political parties across 23 countries, providing a much-needed cross-national empirical evidence on the role echo chambers play in concentrating and isolating extreme views in a political communicative setting. Another important work is Efe Sevin's working paper on how international actors and foreign policy practitioners use digital media to expedite and re-negotiate existing diplomatic processes.[75] Building upon a Twitter-scraped dataset of embassy and consulate connections across the global, Sevin makes the case that middle powers may have disproportionately more significant weight in international diplomacy by seizing upon the amplifying potential of social media. Finally, Caiani and Wagemann demonstrate how the Italian and German extreme far right connect in digital space, exploring aspects of communicative radicalization and network capital of extreme political ideologies.[76] The authors discover that extremist networks cluster and connect differently across political cultures, with separate layers of connectors, leaders and marginalized sub-groups.

**My Personal History with ComInt**

I come from a qualitative background. Given University of Essex being a stronghold of discourse analysis, I developed keen interest in Foucauldian approaches to power and politics. Through my dissertation however, the amount of data I collected on discursive construction of violence, terrorism and conflict became so large that I was unable to deal with them meaningfully through qualitative analysis alone. When I bounced the idea of quantifying discourse with my PhD supervisor, he momentarily panicked, as the practice wasn't as commonplace as it is nowadays. '*You will either get kicked out of the PhD program, or get an award*' was his reply. In the following months, I learned statistics from scratch, engaging in successive crash courses in regression analysis and mathematical modelling offered at the university. These didn't help, as such courses were still taught for extremely large classes with economy, management and political science students with different quantitative skills all pitted into the same class. I learned statistics and regression analysis mostly through self-study (Youtube didn't exist back then).

My resultant dissertation combined a ten-year content analysis of open floor debates in three parliaments, coded and sorted according to sentiment, syntax and lexicon, with another matrix of coding for politicians' ideologies and political interests. Eventually, I demonstrated that regardless of country and political system, conservative and liberal politicians used the same linguistic and sentiment characteristics to define intra-state conflicts. A conservative politician in the United States, Belgium and Turkey sounded significantly more alike, compared to liberal politicians in their own countries, and vice versa. And I could reliably demonstrate the relationship from a statistical point of view. This was good evidence that contributed to the trans-nationalization of conflict behavior and how crisis periods end up internationalizing certain ideologies. Thankfully, my advisor's second prediction ended up happening. I wasn't kicked out of the program and won the Middle East Studies Association's Malcolm H. Kerr dissertation award. Although the methodology isn't new anymore at this point, back when I submitted, quantifying and measuring discourse numerically was considered as a methodological heresy of sorts. This was effectively merging positivist and post-positivist traditions and like one of my examiners put: 'like writing a Muslim Bible'.

The second turning point in my multi-method odyssey was in 2015. After writing my book, I was focusing on the study of armed non-state actor behavior in northern Syria and northern Iraq, both getting increasingly more complex and frustrating to monitor. As actors on the ground quickly exchanged territory, merged, broke-away and disappeared on the battlefield, generating new knowledge or testing theories on conflict were all getting increasingly more difficult. I began generating elementary maps for my own study purposes using Google Earth layers and basic image processing software like Paint. It was around this time that I began to realize that the sophistication of one's maps aren't as important as the story those maps are telling. One of my completely low-tech maps would later be solicited for publication by the New York Times along an op-ed on armed violence in northern Iraq. But getting battlefield information was proving extremely difficult, as the majority of conflict events were taking place across the inaccessible parts of Syria and Iraq. Lucky for me - and perhaps for all conflict scholars - that the advent of mass social media, smartphone and digital propaganda coincided with the war against ISIS. This allowed conflict researchers to extract and process enormous

---

*Review E* 70, no. 6 (December 6, 2004): 66111, https://doi.org/10.1103/PhysRevE.70.066111.
[74] Jonathan Bright, "Explaining the Emergence of Echo Chambers on Social Media: The Role of Ideology and Extremism," *arXiv:1609.05003 [Physics]*, September 16, 2016, http://arxiv.org/abs/1609.05003.

[75] Efe Sevin, "Traditional Meets Digital: Diplomatic Processes on Social Media" (International Studies Association Annual Conference, Baltimore, MD, 2017).
[76] Manuela Caiani and Claudius Wagemann, "Online Networks of the Italian and German Extreme Right," *Information, Communication & Society* 12, no. 1 (February 1, 2009): 66–109, https://doi.org/10.1080/13691180802158482.

volumes of digital content shared by the locals, citizen journalists and the militants, who documented war deep inside the fog of war. A young militant for example, could share how their group struck a Syrian Army tank and post it online with a video and associated hashtag, intended as propaganda, but ending up becoming a data node for conflict researchers. I began scraping some of that content through Twitter, Instagram (before they changed their API) and Flickr. Perhaps the most interesting detail about these battlefield posts were the selfies, that constituted a large portion geotagged conflict events, which practically exposed them on the battlefield. I continued to scrape more of such tweets, eventually coming up with a test run of around 17,352 geotagged tweets through a two-year period, mapping them through time-frequency diffusion. The resultant combination of those images and videos gave me one of the most detailed and high-granularity war map that was out there at the time, rivaling the level of detail of many state-produced war maps that exited at the time, which I published with the Financial Times and Journal of International Affairs.

I took my newfound methodological odyssey to Oxford, where I went as a visiting fellow at the Oxford Internet Institute (OII). OII was an interdisciplinary Oxford department, dedicated solely to the study of the Internet and digital data, along with its political, economic, social and psychological effects on human relations. It was made up of a very uncommon combination of scholars, from physics, biology, geography, computer science, mathematics and political science, all trying to approach different theoretical topics related to the Internet, through a multitude of methodologies. It was there that I learned how Gaussian particle physics principles could help explain how people chose their mates on Tinder and other online dating platforms, or how epidemiological models in biology could explain how riots and protests emerge and spread. It was there that I learned how to code (thanks to computer science doctoral training program for having me audit their Python classes), conduct network measurement, build algorithms for text mining, use more advanced mapping and network analysis software to dig deeper into the logic of large coding structures. I was then admitted to the Alan Turing Institute in London, where I had a chance to participate in data science research groups from Cambridge, Warwick, UCL and Edinburgh, that focused on urban analytics, extremism networks and measuring human digital behavior.

The luxury of months of incubation and daily access to some of the brightest and pioneering minds on computational studies forced me to think about the future of IR, its methodological debates and how computational tools can be incorporated into the study of world events. My first project was an expanded and improved version of the earlier work on battlefield data. Focusing solely on selfies, I scraped battlefield digital data from geographically confined locations in Syria, using a corpus of keywords in English, Farsi, Kurdish and Arabic and generated an event dataset that only contained armed incidents. In my second project, I developed my earlier work on measuring how pre-digital and digital forms of mobilization influenced protest and resistance networks, by focusing on Turkey's failed coup on 15 July 2016. Then, I took part in multiple research clusters on how cultural and geographical proximity between cities helped us measure the likelihood of conflict, how dyadic and multi-level sentiment analysis of digital text allow us to predict radicalization and terrorism networks, and using machine learning algorithms to visually detect and predict the likelihood of armed conflict using Google OpenMap images.

The culmination of my dual visit to Oxford Internet Institute and the Alan Turing Institute was my 'Turing Lecture' in the latter one, where I made an introduction and exposition of the term 'Computational IR', detailing how data science and international relations can form a productive partnership, in a way that doesn't only benefit these two disciplines, but also form as the basis of collaboration with the full range of natural sciences scholars and social scientists. Since - to the best of my knowledge - there has been no previous use of the term, I'd like to coin 'Computational International Relations' as a way to establish a new methodological field that hopefully will transcend the traditional quantitative-qualitative schism in the field, as well as in social sciences.

*Method Training: Or 'How to be a Computational IR Scholar'*

Up until very recently, ComSoc training existed largely within the confines of quantitative-leaning social scientists taking data science courses, with the exception of the Oxford Internet Institute, Harvard Institute for Quantitative Social Science and Stanford University's Computational Social Science Program, which are pioneering institutions of the field. Currently there are a wide range of choices for social scientists from summer courses to master's degrees dedicated to computational social science. To the best of my knowledge, there are no ComInt programs; rather, IR scholars currently can take ComSoc courses and create their own sub-specialization. Oxford's Center for Technology and Global Affairs, where I'm currently a research fellow, is also gearing up to fill in this vacuum in the near future.

There are two aspects of ComInt training. The first one is the easier part: what kind of a technical foundation should students develop? Different ComSoc programs provide different curricula for this purpose, but there are common denominators. Data visualization, model construction and estimation, along with honing statistical skills is generally the first step. Later, understanding different data types used in computing, and various processing principles – clustering, event-driven simulation, approximating functions, derivatives and basic Monte Carlo techniques - are required to build upon the initial foundation. At this time, introductory knowledge of Java, Shiny, Python, R and C++ should be introduced, along with mainstream programs such as ArcGis or QGis (for geospatial analysis), NVivo, RapidMiner or QDA (for text analysis), Gephi, iGraph or NodeXL (for network analysis), and Repast, Swarm, EpiModel or MASON (for various modelling analyses). These technical skills must be reinforced through qualitative/historical theoretical courses on spatial analysis, complexity research, logic of algorithms and basic neuroscience (mostly for complexity research). Final touches can be made through large dataset maintenance skills through Entity-Relationship Diagram (ERD), SQL (Structured Query Language), data definition language (DDL) and data manipulation language (DML).

The second aspect of ComInt/ComSoc training is the harder part: understanding how much technical skill you need to learn and sustain for your own research career. Like any language, computer science requires sustained daily use to remember and preserve knowledge. To that end, being a ComInt/ComSoc scholar means a) knowing you can't master all computer science tools, b) balancing between the main task of social scientists (theory-building) and methods-driven nature of computer science and c) understanding which computational tools you need to develop and which ones to outsource. My answer to all three questions – at least for graduate students – is: be promiscuous. Spend at least six months to dig deep into the computer science world and immerse yourself in methods-driven research. Build your R packages, learn how to scrape Twitter data and spend some time visualizing them on a multitude of spatial, network and text-based software. Although most senior social scientists will advise you to not forget the fact that you are a social scientist, my advice is: forget it – at least for a limited period of time. This period is critical to learn how to think like a computer scientist; not just to get a new perspective, but also to understand the basics of computer-driven research. This is crucial, as although computer science methods are constantly evolving, basic principles of computers (automation, the logic of repeating work, strings, data structures, loops, variables, functions etc.) don't change radically. You can quickly adapt to new programming languages and platforms, once you know what a programming language does. It is also only after spending several months on coding that students can get a sense of what computational tools can do for their own research agenda and the kind of questions they seek to ask.

The final part of ComInt/ComSoc training is the hardest part: forget everything. This phase is about deliberately stopping computer science work and return back to IR, PolSci or another social science discipline of origin. My advice would be to re-read the core theoretical readings of the field the student is coming from and to rethink the fundamentals of the field following several months of immersion in the world of programming. Another twist to this suggestion would be to return back to another social science discipline, instead of the student's own point of origin. To give an example, an IR student going through the computational curve should ideally go to sociology and history to establish an introductory foundation there, creating a triangular expertise. Although not easy, mastery of IR and good introductory knowledge of computer science, and sociology or history will expand the student's analytical prowess significantly.

**Case Studies from My Research Trajectory: How Can Conflict Researchers Benefit from ComInt?**

ComInt is hard to explain by demonstrating its application on one single research question. One significant line of research in ComInt focuses on the relationship between social media and international or comparative political processes under conflict. Some of the most important works in this field are: Tucker et. al. (2017) work on the relationship between social media and democracy,[77] Steinert-Threlkeld's study on the effects of Internet on social mobilization,[78] Rød and Weidmann's work on comparative authoritarianism and the Internet,[79] Anita Gohdes' study on how regimes hide their atrocities on the Internet,[80] (as well as her important overview of how the use of Internet data has changed the study of conflict[81]), Mitts' work on ISIS radicalization on Twitter,[82] Little's formal modelling work on how ICTs affect protest behavior,[83] Zeitzoff's review of how social media is changing conflict[84] (and his social experiment of the 2012 Gaza conflict[85]) and Gunitsky's work on how autocracies use social media as a form of regime stabilization tool.[86] The list is far from complete, however, as the discipline and its exciting methods are rapidly evolving and improving.

---

[77] Joshua A. Tucker et al., "From Liberation to Turmoil: Social Media And Democracy," *Journal of Democracy* 28, no. 4 (October 7, 2017): 46–59, https://doi.org/10.1353/jod.2017.0064.
[78] Zachary C. Steinert-Threlkeld, "Spontaneous Collective Action: Peripheral Mobilization During the Arab Spring," *American Political Science Review* 111, no. 2 (May 2017): 379–403, https://doi.org/10.1017/S0003055416000769.
[79] Espen Geelmuyden Rød and Nils B. Weidmann, "Empowering Activists or Autocrats? The Internet in Authoritarian Regimes," *Journal of Peace Research* 52, no. 3 (May 1, 2015): 338–51, https://doi.org/10.1177/0022343314555782.
[80] Anita R. Gohdes, "Pulling the Plug: Network Disruptions and Violence in Civil Conflict," *Journal of Peace Research* 52, no. 3 (May 1, 2015): 352–67, https://doi.org/10.1177/0022343314551398.
[81] Anita R. Gohdes, "Studying the Internet and Violent Conflict," *Conflict Management and Peace Science*, October 25, 2017, 738894217733878, https://doi.org/10.1177/0738894217733878.
[82] Tamar Mitts, "From Isolation to Radicalization: Anti-Muslim Hostility and Support for ISIS in the West," SSRN Scholarly Paper (Rochester, NY: Social Science Research Network, March 31, 2017), https://papers.ssrn.com/abstract=2795660.

Here, I'll steer clear of engaging in yet another literature review and instead, try to explain how computational tools improved my own scholarship across different topics in IR by adopting hybrid methods. My first exposure to computational research has been through the nudge of a group of recently graduated Harvard computer science PhDs, who wanted to tackle issues related to security and conflict. IR students will receive similar calls from computer scientists. If not, they should initiate contact themselves, either through speaking to a computer scientist faculty, or peers in the computer/data science department. Such calls are usually the first step for any social scientist to collaborate with computer scientists. However, interdisciplinary research builds its momentum slowly, can be frustrating and this shouldn't discourage new researchers from being persistent and continuing to engage with research partners. Although my research partners decided to set up a startup and drifted away from our research eventually, I learned the basics of web scraping, setting up web crawlers and using API data from them. These tools would ultimately be instrumental in my research project on mapping militant selfies in Syria.

Conflict research in IR has developed a keen interest in event data in recent years. From statistical to geographic layers, event data enables us to track conflict patterns, targeting choices and border contestations across a single, or multiple conflict settings. But the majority of that event data (such as UCDP/PRIO or UMD) comes from 'official' sources, derived largely from state-level resources, of mainstream media companies that report on battlefield developments. But what about the inaccessible parts of a conflict? What if neither reporters, nor intelligence operatives of state actors can access no-go zones in a conflict and how do we get event data from there? Up until 2012-13, a clear answer was hard to provide. Thankfully for researchers, non-state actors' use of digital technologies, smartphones and social media have led to a strange setting where active combat and insider developments in no-go zones are broadcast digitally on a minute-by-minute basis with geotags. Militants overwhelmingly began using social media to publicize important events such as armed clashes, declarations of loyalty, or reports of death. Most groups in Syria such as ISIS, YPG or FSA have learned to catalogue these events online with dedicated hashtags and visuals for propaganda, making sure that they are easily searchable. For the exact same reason, they make excellent computational conflict data. In the first phase of my research, I have scraped around 15,000 selfies from Syria, all belonging to Kurdish groups YPG, SDF and their offshoots, through January 2014 - June 2016. Building a word corpus consisting of words related to armed events (bombing, shooting, explosion, airstrike...) I've applied entity-recognition algorithm to scrape all tweets containing these keywords in pre-set coordinates isolating northern Syria, and containing photos that were taken with the front camera of a smartphone (back then, this was the best way of scraping selfie data). I mapped out the resultant dataset to infer where Kurdish groups were fighting, where they were defending and which battles were they avoiding.[87] This became the foundation of my article on Kurdish geopolitics later on.[88]

While I was planning to expand the militant selfie study to other groups in Syria and also bring in Ukrainian groups, a failed coup attempt took place in Turkey, in July 2016. I reorganized my work to focus on the digital engagement patterns during the coup attempt and started to scrape geotagged tweets that clustered around six most widely shared hashtags. These hashtags not only gave me which districts mobilized the most against the coup attempt, but also generated a valuable dataset to model later on through physics or epidemiological approaches. Several things stood out from the study: first, it was religious networks (tariqas), rather than political party networks of AKP that had initiated the first mobilization against the coup. Although AKP networks later mobilized to significantly increase the numbers in the streets, tariqa-dominant districts have been deployed faster and at greater volume during the earlier hours of the coup attempt.[89] This computational data is important

---

[83] Andrew T. Little, "Communication Technology and Protest," *The Journal of Politics* 78, no. 1 (December 17, 2015): 152–66, https://doi.org/10.1086/683187.
[84] Thomas Zeitzoff, "How Social Media Is Changing Conflict," *Journal of Conflict Resolution* 61, no. 9 (October 1, 2017): 1970–91, https://doi.org/10.1177/0022002717721392.
[85] Thomas Zeitzoff, "Does Social Media Influence Conflict? Evidence from the 2012 Gaza Conflict," *Journal of Conflict Resolution* 62, no. 1 (January 1, 2018): 29–63, https://doi.org/10.1177/0022002716650925.
[86] Seva Gunitsky, "Corrupting the Cyber-Commons: Social Media as a Tool of Autocratic Stability," *Perspectives on Politics* 13, no. 1 (March 2015): 42–54, https://doi.org/10.1017/S1537592714003120.
[87] H. Akın Unver, "Mapping Militant Selfies: Application of Entity Recognition/Extraction Methods to Generate Battlefield Data in Northern Syria" (Computer Science Doctoral Training Program Seminar, Oxford University, May 31, 2017), https://www.cs.ox.ac.uk/seminars/1855.html.
[88] H. Akin Unver, "Schrödinger's Kurds: Transnational Kurdish Geopolitics In The Age Of Shifting Borders," *Journal of International Affairs* 69, no. 2 (Spring/Summer 2016): 65–98.
[89] H. Akin Unver and Hassan Alassaad, "How Turks Mobilized Against the Coup," *Foreign Affairs*, September 14,

because it gives us a good early measurement of digital sociology: in an increasingly interconnected world, what is the most foundational source of collective action? During emergencies and times of uncertainty, which fundamental social organizing forces manages to generate the momentum enough to mobilize masses into collective action? The case of Turkey's failed coup reveals to us that at least in Turkish socio-cultural case, religious networks fill in this emergency role. I'm currently at the point of expanding this study into how different religious movements adapt to digital technologies and generate collective action in the US, Hungary, Serbia, Ukraine and Israel.

Simultaneously, I've been drawn further into the concept of digital spoilers and distractors. Bot research is a growing and popular area of study, yet we still know so little about their role in international relations and how they influence global crises. My conversations with Phil Howard, the director of computational propaganda project, reinforced my view that much of the research on bots is dedicated to their impact in politics and sociology, but not enough on IR. To that end, I've begun collecting real-time data during particular international crises to measure anomalies in hashtags and fake news diffusion. My hypothesis, based on raw personal observation during digital crises was that bot-driven hashtags were more likely to disproportionately increase during very short periods (15-20 minutes) and end abruptly, without organic sustain. In contrast, organically-driven hashtags, usually increase more gradually and are sustained over the course of several hours, sometimes days. The first incident I could test this hypothesis was the Saudi Arabia, Qatar and UAE diplomatic crisis, which began in June 2017, as a response to statements attributed to the Qatari Sheikh. A time-frequency geospatial analysis of the most frequently shared hashtags show us that the majority of anti-Qatari messaging were driven by bots.[90] This finding was important because it was one of the first measureable evidence that countries use social media to escalate, signal and pressure other countries into desired behavior. By driving mass anti-Qatar hashtags on social media Saudi Arabia and UAE were using a new way of diplomatically pressuring Qatari leadership to toe the line. A second case where I could further test my hypothesis was the Al Aqsa riots in July 2017. Measuring diffusion patterns of seven widest-shared hashtags, I was able to infer bot-driven versus organically-driven messaging, giving me a good idea on external countries trying to influence foreign crises.[91] This gives us good data to test and challenge some of the central IR hypotheses such as signaling, bargaining, pressuring and diversionary conflict theories.

Currently, I'm a co-principal investigator in a Turing-funded study that aims to build an artificial-intelligence based conflict event detection database. The project combines some of the methodological perspectives I've discussed here – text mining, network analysis, geospatial data – to automatically harvest battlefield digital data in order to generate armed events and log them in real-time (with some redundancy-check lag, of course).

**Conclusion - How Can IR Benefit from Big Data and Machine Learning?**

Big data and computational approaches to social research has revolutionized social sciences and will inevitably impact how IR methods evolve in the coming decades. Two factors define the potential of computational research; sheer size of data that is extremely hard (often impossible) to process with conventional tools of quantitative or qualitative analysis and the advent of more powerful tools that allow us to zoom in and out of various levels of human behavior simultaneously. From this perspective alone, the big data revolution will force us to rethink a fundamental component of IR research: the levels of analysis problem. Big data gives us data granularity that enables micro-level approaches such as behavior, cognitive biases or worldview analysis, as well as the volume that can be scaled to meso-level (networks, collective action, ethno-nationalist movements) and macro-level (ideology, identity, systems research) simultaneously. When done properly, big data and computational tools allow us to model and understand human behavior much better than past approaches, while it is also easier to misuse and exaggerate their explanatory power.

One of the main problems with big data research is an over-reliance on the processing power of the tools, without an eye on cultural and local differences in data. One very common line of dreadful mistake I encounter is usually in social media extremism research that concerns jihadi networks. When engineer or programmer-dominated research groups employ computational tools in extremism research without a social scientist, and/or a scholar with area and cultural expertise, they overwhelmingly produce faulty machine learning word corpus clusters. These clusters often confuse religious statements that express radical behavior with commonplace, regular cultural religious expressions. One computer science conference paper I've had the misfortune of reading (and won't cite) had built a corpus of jihadi radical word corpus, which included common religious terms that Muslims use

---

everyday, such as 'Allāhu akbar' or 'InShaAllah', fundamentally skewing the results. Although this was an extreme case of tone deafness in computational research, there are very frequent, common and subtler ways of bias in research that is produced by research groups that aim to tackle culturally-sensitive social science research. Equally problematic are social science research clusters that aim to build machine learning algorithms by checking Youtube tutorials, without using a computer science specialist, or generate behavioral models without a dedicated modeller. Computational research reaches its true potential in truly multi-disciplinary research clusters and this is precisely why facilitator networks that bridge diverse sciences disciplines and establish a common language across them is the most urgent and important step universities must take in initiating computational research groups.

That is why machine learning, as a way of enabling computers to build new ways of approaching evolving tasks, without being explicitly programmed to solve them, is a field that should go beyond computer science and needs the attention of social scientists. Since we can (and in the near future, will) build machine learning algorithms to track the extent of nationalist sentiment in multiple countries, explore real-time public opinion during an international crisis, or how armed or non-armed non-state actors behave during a violent conflict, the topic doesn't fall far off of the radar of international relations. Not only should future IR doctoral students and early career academics will encounter issues related to big data, computational social science and machine learning, some of them will have to build a foundation in reading and understanding how algorithms work and how to communicate with computer scientists for collaborative research. This means that IR PhDs will have to learn Python or R as a foundational programming language, and add a second software (like ArcGis, Gephi, LingPipe, Ontotext) that fits their immediate research needs.